\documentclass[preprint,showpacs,preprintnumbers,amsmath,amssymb]{revtex4}

\newcommand\rf[1]{(\ref{eq:#1})}
\newcommand\lab[1]{\label{eq:#1}}

\newcommand\br{\begin{eqnarray}}
\newcommand\er{\end{eqnarray}}
\newcommand\be{\begin{equation}}
\newcommand\ee{\end{equation}}

\newcommand\lb{\lbrack}
\newcommand\rb{\rbrack}

\newcommand\bc{\begin{center}}
\newcommand\ec{\end{center}}

\newcommand\pa{\partial}

\newcommand{\ct}[1]{\cite{#1}}
\newcommand{\bib}[1]{\bibitem{#1}}
\newcommand\PRL[3]{\textsl{Phys. Rev. Lett.} \textbf{#1}, #3 (#2)}

\begin{document}

\preprint{arxiv:[hep-th]}

\title {Localized Axion Photon States in a Strong Magnetic Field }

\author{E.I. Guendelman}%
\email{guendel@bgumail.bgu.ac.il}
\affiliation{%
Department of Physics, Ben-Gurion University of the Negev \\
P.O.Box 653, IL-84105 ~Beer-Sheva, Israel
}%

\begin{abstract}
We consider the axion field and electromagnetic waves with rapid time dependence, coupled to a strong time independent,
asymptotically approaching a constant at infinity "mean" magnetic field, which takes into account the back reaction 
from the axion field and electromagnetic waves with rapid time dependence  in a time averaged way. The direction of the self consistent mean field
is orthogonal to the common direction of propagation of the axion and electromagnetic waves with rapid time dependence and parallel to the polarization of these electromagnetic waves. Then, there is an effective U(1) symmetry mixing axions and photons. 
Using the natural complex variables that this U(1) symmetry suggests we find localized planar soliton solutions. These solutions appear to be stable since they produce a different magnetic flux than the state with only a constant magnetic field, which we take as our
"ground state". The solitons also have non trivial $U(1)$ charge defined before, different from the uncharged vacuum.
\end{abstract}


\maketitle

\section{Introduction}

One of the most interesting ideas for going beyond the standard model
has been the introduction of the axion \ct{weinberg}, which provided a way to solve the strong CP problem.
Since then, the axion has been postulated also as a candidate for the dark matter.
A great number of ideas and experiments for the search this particle have been proposed
\ct{review}.

One particular feature of the axion field $\phi$ is its coupling to the photon 
through an interaction term of the form 
$g \phi \epsilon^{\mu \nu \alpha \beta}F_{\mu \nu} F_{\alpha \beta}$.
In fact a coupling of this sort is natural for any pseudoscalar interacting with
electromagnetism, as is the case of the neutral pion coupling to photons (which
as a consequence of this interaction decays into two photons).

A way to explore for observable consequences of the coupling of a light scalar 
to the photon in this way is to subject a beam of photons to a very strong magnetic field.

This affects the optical properties of light which could lead to testable consequences\ct{PVLAS}. 
Also, the produced axions could be
responsible for the "light shining through a wall phenomena ", which  is obtained by first
producing axions out of photons in a strong magnetic field region, then subjecting the mixed beam of photons
and axions to an absorbing wall for photons, but almost totally transparent to axions due to their weak 
interacting properties which can then
go through behind this "wall", applying then another magnetic field one can recover once again some photons 
from the produced axions \ct{LSTW}.

On the other hand, the photon axion system in an external magnetic field shows very interesting field theoretical 
properties, like the possibility of mass generation \ct{Ansoldi} and a new possibility for confinement \ct {confinement}.
In particular the results of \ct{Ansoldi} imply that an external constant magnetic field in the axion photon system
is a stable vacuum under small perturbations. This is in contrast to an external electric field, which exhibits 
tachyonic instabilities \ct{Ansoldi}.

Here we will consider the effects which result from taking into account the back reaction of the axions and electromagnetic waves
on this strong "external" magnetic field.  It will be shown that
when considering the axion photon system in an self consistent, time independent magnetic field, 
that is one that takes into account the back reaction of the axion photon system on the magnetic field
in an averaged mean field approach,
we find that there are localized soliton like solutions. 
The direction of the self consistent mean field we will take
 orthogonal  to the direction of propagation of the axion and propagation of the electromagnetic component with rapid time dependence
and parallel to the polarization of the electromagnetic waves. 
The resulting planar soliton solutions that are obtained appear to be 
stable since they produce a different magnetic flux than configuration where the localized axion photon 
configuration is absent, that is, when we have just the a constant magnetic field all over space (our choice of stable vacuum).

\section{Action and Equations of Motion}
The action principle describing the relevant light pseudoscalar coupling to the electromagnetic field is,
\be
S =  \int d^{4}x 
\Bigl\lb -\frac{1}{4}F^{\mu\nu}F_{\mu\nu} + \frac{1}{2}\pa_{\mu}\phi \pa^{\mu}\phi - 
\frac{1}{2}m^{2}\phi^{2} - 
\frac{g}{8} \phi \epsilon^{\mu \nu \alpha \beta}F_{\mu \nu} F_{\alpha \beta}\Bigr\rb
\lab{axion photon ac}
\ee

We now specialize to the case where we consider a time dependent axion and electromagnetic field with propagation only along the $z$-direction
and where a time independent magnetic field pointing in the $x$-direction is present. This field may have a $z$ dependence, 
to be determined later, but it is taken to be time independent. We want however that this field will take into account the back reaction of the time dependent axion and electromagnetic fields
in a time averaged way. In the case the magnetic field is constant and external, see for example \ct{Ansoldi} for general solutions.

Now considering a static strong "mean" magnetic field pointing in the x direction having an
arbitrary z dependence, we take $B_x = -\pa_{z} A_y$ (since taking also that $ A_z$ depends only on z leave us only with this contribution) and specializing to z dependent electromagnetic field perturbations and axion fields. This means that the interaction between the strong mean field, the axion and photon fields (that is the part of the electromagnetic field with fast variations)
reduces to
 
\be
S_I =  -\int d^{4}x 
\Bigl\lb  gB_x(z)\phi E_x\Bigr\rb
\lab{axion photon int}
\ee

Choosing the temporal gauge for the electromagnetic field and considering only the x-polarization for the electromagnetic time dependent fields ($A$ will denote the x-component of the vector potential, so $ E_x = -\pa_{t} A$), since only this polarization couples to the axion and static magnetic field. We get the following 1+1 effective dimensional action,

\be
S_2 =  \int dzdt 
\Bigl\lb  \frac{1}{2}\pa_{\mu}A \pa^{\mu}A+ \frac{1}{2}\pa_{\mu}\phi \pa^{\mu}\phi - 
\frac{1}{2}m^{2}\phi^{2} + gB_x(z) \phi \pa_{t} A - \frac{1}{2}(\pa_{z} A_y)^{2}
\Bigr\rb
\lab{2 action}
\ee

($A=A(t,z)$, $\phi =\phi(t,z)$), which leads to the equations for the $A$ and $\phi$ fields, if we assume the
mean field $B_x(z)$ is time independent

\be
\pa_{\mu}\pa^{\mu}\phi + m^{2}\phi =  gB_x(z) \pa_{t} A
\lab{eq. ax}
\ee

\be
\pa_{\mu} \pa^{\mu}A = - gB_x(z) \pa_{t}\phi 
\lab{eq. photon}
\ee

The eq. of motion of the mean field $ A_y(z) $ itself will be discussed later.
As it is known, in temporal gauge, the action principle cannot reproduce the Gauss 
constraint  and has
to be impossed as a complementary condition. However this 
constraint is automatically satisfied here just because of the type of dynamical reduction
employed (which gives both that $\pa_{i}E^{i}=0$ and that the charge density that is proportional to $\pa_{i}(\phi B^{i})= 0$ also)
and does not need to be considered  anymore.

\section{The Continuous Axion Photon Duality Symmetry in the Massless Axion Case and Static Strong Mean Field}
Without assuming any particular z-dependence for $B_x(z)$, but still insisting that
this "mean field"  will be static, we see that in the case $m=0$, we discover a continuous axion 
photon duality symmetry, as we discussed in Reference \ct{Guendelman}, where the magnetic field in the $x$ direction
was taken as external, but we only need the time independence of this mean field to observe that,

1. The kinetic terms of the photon and
axion allow for a rotational $O(2)$ symmetry in the axion-photon field space.

2. The interaction term, after dropping  a total time derivative (if the mean field is static) can also be expressed in 
an $O(2)$ symmetric way as follows

\be
S_I =  \frac{1}{2} \int dzdt 
gB_x(z) \Bigl\lb \phi \pa_{t} A - A \pa_{t}\phi \Bigr\rb
\lab{axion photon int2}
\ee

The axion photon symmetry is in the infinitesimal limit
\be
\delta A = \epsilon \phi, \delta \phi = - \epsilon A
\lab{axion photon symm}
\ee

where $\epsilon$ is a small number. Using Noether`s theorem, this leads to the 
conserved current $j_{\mu}$, with components given by

\be
j_{0} = A \pa_{t}\phi - \phi \pa_{t} A + \frac{gB_x(z)}{2}(A^{2} + \phi^{2} )
\lab{axion photon density}
\ee
and 
\be
j_{i} = A \pa_{i}\phi - \phi \pa_{i} A 
\lab{axion photon current}
\ee
defining the complex field $\psi$ as
\be
\psi = \frac{1}{\sqrt{2}}(\phi + iA)
\lab{axion photon complex}
\ee
we see that 
in terms of this complex field, the axion photon density takes the form
\be
j_{0} = i( \psi^{*}\pa_{t}\psi - \psi \pa_{t} \psi^{*}) +  gB_x(z) \psi^{*}\psi 
\lab{axion photon density complex}
\ee

We observe that to first order in $gB_x(z)$, \rf{axion photon int2} represents the
interaction of the magnetic field with the "axion photon density" \rf{axion photon density}, \rf{axion photon density complex} 
and also this interaction has the same form as that of scalar QED with an external "electric " field to first order.
In fact the magnetic field or more precisely $gB_x(z) /2$ appears to play the role of external electric potential.

In terms of the complex field, the axion photon current takes the form
\be
j_{k} = i( \psi^{*}\pa_{k}\psi - \psi \pa_{k} \psi^{*}) 
\lab{axion photon current complex}
\ee

\section{The Localized Soliton Solutions}
After the manipulations of the previous section, valid under the approximation that the mean field $ A_y(z) $ is time
independent, we can discuss the eq. of motion for this mean field.
This is,

\be
\pa_{z}(\frac{ig}{2}( \psi^{*}\pa_{t}\psi - \psi \pa_{t} \psi^{*}) + B_x(z)) = 0
\lab{B eq.}
\ee

The same result can be obtained from the original equations instead of the averaged Lagrangian obtained under the assumption that the mean field $B_x(z)$ is time independent and there doing a time averaging procedure, using for example
that under such time averaging $\phi \pa_{t} A$ equals $\frac{1}{2}(\phi \pa_{t} A -\pa_{t}\phi A)$ .
Equation \rf{B eq.} can be integrated, giving

\be
B_x(z) = - \frac{ig}{2}( \psi^{*}\pa_{t}\psi - \psi \pa_{t} \psi^{*}) + B_0
\lab{mean field solution}
\ee

where $B_0$ is an integration constant. The constant $B_0$ breaks in fact spontaneously the charge conjugation symmetry of the theory,

\be
\psi \rightarrow \psi^{*}  
\lab{charge conjugation}
\ee

or equivalently, changing the sign of $A$, since in such transformation the first term of \rf{mean field solution} changes sign, which would be required to leave the  interaction term \rf{axion photon int2} in the action invariant,
but the second term does not (since it is a constant).
Also, in problems where $B_x$ is taken as an external field \ct{Guendelman2}, the interaction automatically breaks this "charge conjugation"
symmetry.

We now consider $\psi $ to have the following time dependence,

\be
\psi = exp(-i \omega t) \rho(z)
\lab{time dep.}
\ee

We want to see now what is the equation of motion for $\rho(z)$, which we take as a real field. We start with the general eq. for $\psi$

\be
\pa_{\mu}\pa^{\mu} \psi + igB_x(z)\pa_{0}\psi = 0
\lab{complex eq.}
\ee

Inserting \rf{time dep.} into \rf{mean field solution} and the result into \rf{complex eq.}, we obtain,

\be
\frac{d^{2}\rho(z)}{dz^{2}} + \frac{d V_{eff}(\rho)}{d \rho} = 0
\lab{analog mechanical}
\ee

where $V_{eff}(\rho)$ is given by

\be
V_{eff}(\rho) = \frac{1}{2}(\omega^{2} -  \omega g B_0)\rho^{2} + \frac{1}{4}g^{2}\omega^{2}\rho^{4}
\lab{effective potential}
\ee

Some comments are required on the nature and signs of the different terms.
One should notice first of all that this effective potential is totally dynamically generated and vanishes when taking $ \omega = 0 $. 
Concerning signs, all terms proportional to $\omega^{2}$
are positive, in fact although the $(g\omega)^{2}$ term is quartic in $\rho$, it has to be regarded as originating not from  an ordinary potential of the scalar field in the original action, but rather from a term
proportional to a  $g^{2}(- i( \psi^{*}\pa_{t}\psi - \psi \pa_{t} \psi^{*}))^{2}$ , quadratic in time derivatives, which could have been
 obtained if we had worked directly with the action rather than with the equations of motion, replacing \rf{mean field solution} back into the action (i.e., integrating out the $B_x$ field). Such type of quadratic terms in the time derivatives give a positive contribution both in the lagrangian and in the energy density, unlike a standard (not of kinetic origin) potential, where the contribution to the lagrangian is opposite to that of their contribution in the energy density.

The only term which may not be positive is the $-\omega g B_0$ contribution. This term breaks the charge conjugation symmetry
\rf{charge conjugation} which for a field of the form \rf{time dep.} means $\omega \rightarrow -\omega$.

We can in any case choose the sign of $\omega$ such that the $-\omega g B_0$ contribution is negative and choose $ B_0$ big enough
(or  $\omega $ small enough) so that this term makes the first term in the effective potential negative.

Now we are interested in obtaining solutions where 
$B_x(z) \rightarrow B_0$ as $z \rightarrow  \infty$ and also as $z \rightarrow  -\infty$, which requires $\rho \rightarrow 0$ as $z \rightarrow  \infty$ and also as $z \rightarrow  -\infty$. Since the vacuum with only a constant magnetic field is a stable one \ct{Ansoldi}.

The solution of the equations \rf{analog mechanical} and \rf{effective potential} with such boundary conditions is possible if
$\omega^{2}  - \omega g B_0 < 0$ . After solving these analog of the "particle in a potential problem" with zero "energy", so that the boundary conditions are satisfied, we find that $\rho$ is given by (up to a sign),

\be
\rho = \frac{(\sqrt{2(\omega g B_0 - \omega^{2}) })/g\omega}{cosh(\sqrt{\omega g B_0 - \omega^{2} }(z-z_0))}
\lab{the solution}
\ee

where $z_0$ is an integration constant that defines the center of the soliton.

Inserting \rf{the solution} and \rf{time dep.} into the expression for $B_x$ \rf{mean field solution} we find the profile for
$B_x$ as a function of $z$. The difference in flux per unit length (that is ignoring the integration with respect to $y$ in the $yz$ plane) through the $yz$ plane of this solution with respect to the background solution
$B_x=B_0$ is finite amount. Since magnetic flux is conserved, we take this as an indication of the stability of this solution towards  decaying into the $B_x=B_0$ stable "ground state". 

Notice also that the soliton is charged under the $U(1)$ axion photon duality symmetry \rf{axion photon symm} and the vacuum is not, another evidence for the stability of these solitons. However for any given soliton,
there is no "antisoliton", since the condition $\omega^{2} -  \omega g B_0 < 0$ will not be mantained if we reverse the sign of $\omega$.
This is due to the fact that the vacuum of the theory, i.e. $B_x=B_0$ spontaneously breaks the charge conjugation symmetry \rf{charge conjugation}.

There are some similarities, but also some crucial differences with the one dimensional topological solitons considered in Ref. \ct{T.D. Lee},  where also a complex field with a $U(1)$ global symmetry was considered and a time dependence of the form \rf{time dep.}
as well, which means that the soliton is charged under this $U(1)$ global symmetry, as in our case. The difference is in the type of potential
used and the origin of the potential. Here, the complete potential appears when considering non vanishing $\omega$, while in Ref. \ct{T.D. Lee}, 
 there is an original potential and self interaction in the original action and these self interactions appear with a negative 
sign in the effective potential, contrary to our case where the quartic self interaction enters with a positive sign (because of its kinetic origin as we explained before). Finally, in our case there is also a topological aspect as well, absent in Refs. \ct{T.D. Lee}, which is that the magnetic flux of the soliton differs from that of the vacuum.
 
\section{Conclusions}
In the context of a theory of containing a pseudo  scalar particle coupled to an electromagnetic field in the form
$g \phi \epsilon^{\mu \nu \alpha \beta}F_{\mu \nu} F_{\alpha \beta}$ an external constant magnetic field provides a stable vacuum \ct{Ansoldi}
, unlike a constant electric field \ct{Ansoldi}. This one can understand qualitatively since a magnetic field is protected from decay to a lower energy density
configuration because the magnetic field cannot be screened, while this is not the case for a constant electric field.
We  then study a special type of geometry, where all the space dependence is on only one dimension, which we call $z$ and with fast and slow variables. The fast variables being the axion field and the vector potential in the $x$ direction,the slow variable, the magnetic field
in the $x$ direction or the $y$ component of the vector potential, this is our "mean field".
We then consider time averaging in the equations of motion or in the action so that the mean field is taken to be time independent and then considering the limit of zero axion mass we obtain a continuous axion photon "duality symmetry", and conserved quantities associated.

Introducing complex variables makes the structure of the time averaged equations quite manageable and the planar localized (in the $z$ direction) axion photon solutions are then found. From the profile for
$B_x$ as a function of $z$, we see that there is a difference in the flux per unit length (that is ignoring the integration with respect to $y$ in the $yz$ plane) through the $yz$ plane of this solution with respect to the background solution
$B_x=B_0$. Since magnetic flux is conserved, we take this as an indication of the stability of this solution towards  decaying into the $B_x=B_0$ stable "ground state".

In future research it would be interesting to study generalizations of these solutions and look at the possibiliy of localizing not only with respect to one dimension (here $z$) but with respect to two. Another area of future research could be the consideration of a vacuum with charge density, which changes the discussion of solitons \ct{Bekenstein}. Finally one should also consider the effect of a small axion mass on these solutions.

\section*{Acknowledgments}

I would like to thank  the  Department of Physics and Chemistry and in particular the High Energy and Differential Geometry Theory Group of the Southern University of Denmark in Odense, where this work was completed, for great hospitality and support, in particular from Francesco Sannino 
and Roshan Foadi. I also want to thank Niels Kjaer Nielsen for very interesting discussions.


\end{document}